# Observation of the $Sb_2S_3$-type post-post-$GdFeO_3$-perovskite: A model structure for high density $ABX_3$ and $A_2X_3$ phases.

Wilson A. Crichton [1]*, Fabian L. M. Bernal [2], Jérémy Guignard [1], Michael Hanfland [1] and Serena Margadonna [2]*

[1.] ESRF - The European Synchrotron, 71, rue des Martyrs, Grenoble cedex, 38043, France.

[2.] Department of Chemistry and Center for Materials Science and Nanotechnology, University of Oslo, Oslo, Norway.

* Corresponding authors: crichton@esrf.fr (+33(0)476882269) and serena.margadonna@kjemi.uio.no (+47(0)22855847)

**Abstract**

Ten years have passed since the description of the perovskite to post-perovskite transition in MgSiO$_3$ and its' impact on the mineralogy and rheology of the D'' region at the base of the Earth's mantle. Much work has explored the mechanisms operating during this transformation and their influence on seismic response. In parallel, calculations in *ABX*$_3$ systems have identified potential structures for a denser post-post-perovskite phase occurring at higher pressures. However, experiments have yet to elucidate any higher pressure form, beyond the CaIrO$_3$-type. Here we describe the structures and transformations that lead to the crystallisation of a post-post-perovskite of Sb$_2$S$_3$-type in a GdFeO$_3$-type fluoroperovskite at high pressure conditions. The use of single-crystal techniques gives unique access to the relative crystallographic orientations of all polymorphs. We use this information to extend this description to include other calculated and observed forms that are competitive in *ABX*$_3$ and *A*$_2$*X*$_3$ stoichiometries (e.g. Sb$_2$S$_3$, α-Gd$_2$S$_3$, Be$_3$N$_2$,) and provide substantial information on inter-relationships between these structures, which is critical to the interpretation of transition mechanisms, predicting transition sequences, to the expression of directional properties in those transformed structures when assessing their suitability to modelling e.g. their rheological response in deep-Earth and exoplanet interiors.

## Introduction

It has been 10 years since one of the most significant scientific breakthroughs in our understanding of the structure of the deep Earth [1,2,3]. The seismic discontinuity observed at the D" layer is attributed to the transformation of $MgSiO_3$ perovskite (pv, of $GeFeO_3$-type) to its post-perovskite form (ppv, of $CaIrO_3$-type). While much experimental and theoretical work has been devoted to the study of this transition, further advances are impeded by the extreme conditions of temperature and pressure that are needed to reproduce the real conditions at the D" layer. For this reason, studies of the behaviour of related systems, such as the fluoroperovskites and noble-metal oxides systems that also undergo a pv to ppv transition but at much more easily accessible conditions, have been undertaken. It is elementary to establish the phase diagrams, stability and physio-chemical properties of these ppv phases together with possible post-post-perovskite forms [4-16], which can have a direct bearing on our understanding of deep Earth and super-Earth processes. In parallel, other research has made significant ground in extending comparable $A_2X_3$ systems to similar extremes of pressure and temperature; notably in the $Fe_2O_3$, $Al_2O_3$ and lanthanide sesquioxides and sulfide systems [17-26]. Most recently a hexagonal lattice was proposed for Fe-rich phase produced (with ppv) during a decomposition reaction of $(Mg,Fe)SiO_3$ [27], which is further implied to have a structure related to the o$C$ PPV-II proposition [28]. It is however unclear if these lattices and proposed models are commensurate with each-other, with regard to reported diffraction data for the H-phase [29]. While the determination of the crystallographic structures of pv, ppv and allied forms is fundamental, information on how the orientation of their unit cells is affected by their successive transitions is equally valuable. The inheritance of directional and tensor properties that accompany these transformations are pivotal for the full description, interpretation and preservation of textural and rheological response in the deep Earth.

Compounds, such as $Fe_2O_3$, have been observed to transform, with increasing pressure, from corundum-type structures, through the $Rh_2O_3$(II)- to the $CaIrO_3$-type [17]. In the case of $Fe_2O_3$, the high pressure form is characterized by distinct 6 and 8-fold coordinated $Fe^{3+}$ cation sites and a similar situation is also observed in $Al_2O_3$.[18, 19] This structure is shared by all these chemistries and is the highest-pressure single-phase polymorph observed in these and oxide $ABX_3$ systems, such as $MgSiO_3$. It follows that, while neither $Al_2O_3$ nor $Fe_2O_3$ form a pv they do stabilize as ppv and this allows comparisons to be drawn anew, between $ABX_3$ and $A_2X_3$ stoichiometries. $Al_2O_3$ and other sesquioxide structures are then further predicted to transform to other higher density structures; *viz.* the $Sb_2S_3$- and/or the $\alpha$–$Gd_2S_3$-types (the case of $Sc_2O_3$, $In_2O_3$ and $Y_2O_3$) at high pressures [17-26]. The sequence of corundum, $Rh_2O_3$(II), $CaIrO_3$, $Sb_2S_3$ (and/or to α-$Gd_2S_3$) has been predicted to cover a range of pressures of approximately 5 Mbar - pressures well exceeding those available to oxides in the Earth, and to static experimentation. Therefore, a clear link between $ABX_3$ systems and the sesquioxides (or other $A_2X_3$ system) and its' extension, through the common $CaIrO_3$-type, remains an observational enigma. The formation of equivalent structures at these pressures (or at least the accessibility of analogous forms) as $CaIrO_3$- and post-post-perovskites has significant implications for the mutual solubility of $A_2X_3$ and $ABX_3$ at the extremes of terrestrial stability of oxide phases. The search for suitable analogues to extend the systematics of these $A_2X_3$ and $ABX_3$ systems will also overcome the requirement for a "…perovskite compound in which both the 6-fold- and the 8-fold-coordination sites are occupied by the same cation," [30] and offer an unified account of these systems to the highest pressures under which oxides are stable in the Earth's interior, and beyond.

In this study, we provide complete refinements of the different crystal structures observed during compression of $GdFeO_3$- type $NaFeF_3$ and express the crystallographic orientations of the ppv and post-post perovskite $Sb_2S_3$-type structure with respect to the pv form. We interpret the structural sequence through matrix transformations to underscore the transition mechanisms from pv, to ppv and $Sb_2S_3$-type, as well as to the related α-$Gd_2S_3$-type and $Be_3N_2$-type structures. The former is most-commonly observed in high pressure sesquioxide structures [17-26], and the latter is calculated as a further potential post-ppv form in $ABX_3$ [11], but is also known as a polymorphic form of compounds that also show perovskite, bixbyite and $Al_2O_3$ structures. Our work then highlights and extends the similitude between sesquioxide and perovskite-like systems beyond ppv, through their mutual expression as $Sb_2S_3$- and/or α-$Gd_2S_3$-types (which can both related to the $CaIrO_3$-type, but, critically, by different twin mechanisms). As these transformation matrices and the twinning descriptions included herein are exact (for the same volume conditions), they are of significant value in transferring any symmetry or directionally affected tensor properties from one phase to the next as a first-order estimate of how bulk properties are altered at any transition. In addition, given that all post-pv structures yet expected in $ABX_3$ and $A_2X_3$ structure can be related, it is now possible to use a broader base from which to construct an overall systematic for these anticipated high density structures and to establish which and why any particular trend may be exhibited over another. As such these model structures, and the details of their transformations, are fundamental in modelling a wider range of potential phase diagrams and for estimating the rheology (and other physical properties) of deep- and super-Earth environments and offer constraints to the selection of other potential analogues and actual single-phase structures yet to be determined in high density $ABX_3$ and $A_2X_3$ systems.

## Results

**Structural aspects of the transition sequence.** Compression of GdFeO$_3$-type NaFeF$_3$ pv (Fig.1a) has been investigated by single-crystal methods and by large-volume techniques on polycrystalline samples of the same batch. The pv diffraction patterns (Fig. 2) show rapid evolution up to the transition boundary, as evidenced by the continuous increase in separation of the (200), (020) and (002) Bragg reflections (Fig. 2c.) Consequently, the degree of anisotropy increases with pressure, as reflected by the value of the tilting angle that quickly reaches the approximate limiting value [31] of $\phi = 26.9°$ at 9.5 GPa (Table 1; see also [32]).

Through further increase in pressure the ppv NaFeF$_3$ is formed (Table 2, Fig. 1b), between 9.5 and 13.5 GPa, which also exhibits significant anisotropy, with the *a*- and *c*- lengths at approximately 95% of ambient values and *b* having been reduced to 92% at 21.2 GPa (cf. $a = 3.13774(4)$ Å, $b = 10.2087(1)$ Å, $c = 7.4587(1)$ Å [32]). Normalisation of the cell parameters, shows that the *b*-axis stiffens with pressure, such that both $b/a$ and $b/c$ diminish, while $c/a$ increases linearly with pressure, at reduced rates of $b/a:b/c:c/a \approx$ -6:-3:+1. With this pattern of stiffening of *b* relative to the *a-c* plane, the bulk anisotropy decreases with pressure. The same pattern, though more pronounced, is seen in both calculation and experimental studies of NaMgF$_3$ [4,11,12]. This is a consequence of the location of the highly compressible Na-sites, which align parallel to *b* and perpendicular to the corrugation of comparatively rigid edge- and corner-sharing FeF$_6$ octahedra that form stacked sheets in the *a-c* plane. Any *a*-axis lattice compression must require bond length changes, particularly as the edge-sharing F1 sites are fixed at $a/2$, the Fe sites are lattice-bound and Na are located at $(0,y,¼)$ (Fig. 1b). Therefore, the larger part of the compression comes from displacement of F2, by a combination of rotation (the Fe-F-Fe angle increases to ~150°) and bond length change (the longer Na-F distance reduces to 2.80 Å at 21.2 GPa, while the shorter Na-F1 distances reduce to 2.16 Å). As compression results in significant bond-length shortening, it is not unexpected that the more rapid *b*-axis compression will tend to saturate, leading to a consequent reduction in out-of-plane anisotropy with increasing pressure.

Through continued reduction of atom distances and site symmetry the ppv form undergoes a transition to the Sb$_2$S$_3$-type, between 21.2 and 22 GPa (Table 3, Fig. 1c). The occurrence of the new structure was clearly identifiable with loss of the *C* lattice centering and reflections were harvested using a constrained *oP* lattice with *mmm* Laue symmetry. Assessments of the symmetry based on likely subgroups, lattice extinctions and electron density distribution all suggested *Pnma* being most likely. Refinement of a model based on charge-flipping [33], supplemented by additional Fourier difference map cycles, confirmed this (Table 3). The lattice metric (+/- 20%) and $c^5$ Wyckoff sequence uniquely describes the $Sb_2S_3$-type structure, according to the ICSD classification (such a search highlights $In_2O_3$, $Ti_2O_3$ and $U_2S_3$ chemistries [21,22,24]). There is little or no volume change on transition (to the subgroup) and it is completely reversible, with very little hysteresis (~1 GPa). This transition affords an increase in coordination over both Na and Fe sites, through the combination of lattice distortion and reduction in site symmetries from $fc^2a$ to $c^5$. The topotactic relationship between these two structures is evident from the map of observed electron densities (Fig. 3). The pv and ppv forms are both recoverable [32].

**Transition mechanism.** With the aid of the STRUCTURE RELATIONS [34] code of the Bilbao Crystallographic Server [35], we have reconstructed the high symmetry form (post-perovskite in *Cmcm*) from the low symmetry structure ($Sb_2S_3$-type in *Pnma*). Structures in *Cmcm* can transform to those in *Pnma* via a mechanism that would allow for Wyckoff site splitting; so that two $4c_{Pnma}$ sites are formed from one $8f_{Cmcm}$. The correct stoichiometry is retained as other sites have unaffected multiplicities, by becoming 4c sites in *Pnma* (Fig. 4). Inspection of the site symmetries, using SAM[36], identifies that this transition should be detectable by Raman scattering experiments as a significant increase in both active $A_g$ (4 to 10) and $B_g$ (8 to 20) modes is expected. This general *Cmcm* to *Pnma* mechanism is not unusual and a similar group/subgroup description with *f*- to *c*-site splitting can be used, for instance, to describe a transformation from $CaTi_2O_4$- to $CaFe_2O_4$-type; e.g. in $MgAl_2O_4$[37] ($f^2c^2b$ to $c^7$). In $NaFeF_3$, the application of this mechanism, via matrix description (1):

$$\begin{pmatrix} 0 & 1 & 0 & : & 1/4 \\ 0 & 0 & 1 & : & 1/4 \\ 1 & 0 & 0 & : & 1/2 \end{pmatrix} \quad (1)$$

leads, with compression, directly to the formation of a new Fe-F-Fe liaison between adjacent octahedral 'sheets' and an increase in coordination about the Fe site from 6 to 7 (Fig. 1c). This results in shorter, and average, Fe-F distances of 1.98 and 2.08 Å in $Sb_2S_3$-type. Both of these distances are longer than those in the ppv. This is coincident with the displacement and reduction in symmetry of the Fe site (Fig. 5), from the inversion centre at 4a (2/m..) position to 4c (.m.) $_{Pnma}$, along the new bond direction and is visible in the electron density distribution as the kinking of the previously vertically aligned ppv Fe positions (Fig. 3). It is also reflected by the significant reduction of the cell parameter along the former long ppv *b* dimension. There are strictly no atomic displacements in the structure relative to the short $CaIrO_3$-type *a* direction due to the 4c lattice sites retaining their mirror symmetry. The average displacements of the atomic sites are of the order of 0.4 Å, including any lattice distortion. The spontaneous lattice strain is 0.07.

This description of atomic displacements available from the STRUCTURE RELATIONS code is also directly comparable to static calculations of the soft mode displacements in unstable ppv NaMgF$_3$ [11] (Fig. 5). Taken together, the increase in coordination by linking one Fe site to the 7-fold polyhedra of the adjacent layer is expected to stiffen the structure (and reduce anisotropy). It then follows that this will result in a higher incompressibility and, following the same arguments as those presented for NaMgF$_3$ [11], increased acoustic phonon frequencies. At the same time the observed higher coordination will directly lead to the expansion of the cell parameters ($a$ is longer in Sb$_2$S$_3$) and softening of the high optical frequencies[11]. Therefore, a positive Clapyeron slope is expected for this transition, implying that an increase in temperature will favor the crystallization of ppv over the Sb$_2$S$_3$-type structure [11].

**Influence on topology.** The topotactic relationships between pv and ppv structures is of significant importance in determining the mechanisms possible for the transition to take place [14,15], and, to establish in what way any stratification in the Earth's interior through deformation (or transition mechanism) could give rise to observable seismic anisotropy, particularly at D'' [14,15,38]. The intrinsic topotactic relationship between the successive crystalline structures is also evident from our data via the comparison of lattice orientations (the UB matrices). These are derived directly from the distribution of diffraction peaks of each form in reciprocal space - from the same original crystal which has not otherwise moved. This information is a powerful, inherent, feature of the single-crystal technique, which is not available to regular powder diffraction analysis. The following UB matrices are obtained, and correspond to unconstrained volumes of 212.8, 205.5 and 186.36 Å$^3$, for the pv, ppv and Sb$_2$S$_3$ structures:

$$\begin{pmatrix} -0.07867 & -0.01255 & -0.00828 \\ 0.00615 & -0.06473 & 0.02701 \\ -0.01713 & 0.03440 & 0.04780 \end{pmatrix} \quad \text{pv in } Pbnm \text{ setting} \tag{2}$$

$$\begin{pmatrix} -0.01758 & -0.04270 & -0.00779 \\ 0.12001 & -0.00740 & 0.02658 \\ -0.06571 & -0.00286 & 0.05054 \end{pmatrix} \quad \text{post} - \text{perovskite in } Cmcm \tag{3}$$

$$\begin{pmatrix} -0.06656 & -0.00705 & -0.04533 \\ -0.11097 & 0.02402 & 0.01969 \\ 0.04699 & 0.04654 & -0.01722 \end{pmatrix} \quad Sb_2S_3 \text{ type in } bac \text{ setting } (Pmnb) \text{ of spacegroup 62} \tag{4}$$

Inspection of these UB matrices clearly shows that the pv, the ppv and the Sb$_2$S$_3$-type structures share a common axis, and that the common diagonal lattice plane of the two high pressure phases is normal to a principle lattice direction of the original pv. This results in $(001)pv_{Pbnm}$ || $(001)ppv$ || $(010)Sb_2S_{3Pmnb}$ and, for example, the $(100)pv$ is orthogonal to the $(001)$ and $(-110)$ of the ppv and $(010)$ and $(-110)$ of the $Sb_2S_3$ structure (Fig. 6). Actual calculated angles [39] from the UB matrices vary by up to ~1° from ideal values, due to significant pressure-induced differences in the lattice parameters which form part of the UB matrix construction. The angle calculated from the UBs between the $(100)pv$ and $(010)ppv$ is equal to 16.491° (angle *a*, Fig 6a.), entirely consistent with previous observations from TEM measurements, geometrical reconstruction and predictions [14,15]. An equivalent angle between the $(010)pv$ and $(100)Sb_2S_3$ is 19.770° (angle *b*, Fig 6b.), again including a significant proportion of lattice compression. Figure 6c highlights these relationships between the reciprocal lattices of pv, ppv and $Sb_2S_3$-type structures, shown here along $c^*$ of the pv (*Pbnm*), ppv (*Cmcm*) lattices and $b^*$ (*Pmnb*) $Sb_2S_3$, with the green line along to the <010> direction of the pv (*Pbnm*) lattice. It is natural that twinning is expected, with the high coincident lattice points and the high symmetry between principle lattice planes and directions noted above, and indeed twinning is observed.

**Related structures and chemistries.** Given that the high pressure structures in the $Sc_2O_3$, $In_2O_3$ and $Y_2O_3$ chemistries are proposed to share the α-$Gd_2S_3$-type structure [21,24,26], we have also investigated the possibility of generating further twins of α-$Gd_2S_3$-type lattices from ppv lattice points. One possible mechanism that will transform the *Cmcm* ppv lattice, via *cab* and origin shift of (¾¼0) to *Pnma*, generates pairs of *oP* lattices of 5.597(2), 3.016(2) and 11.171(9) Å at 90° to each other about their common $c^*$ (Fig. 6d). These dimensions are consistent with those detailed elsewhere for oxide structures [21, 24, 26]. We tested this against the data and, while symmetry was estimated as *Pnma* (i.e. a lattice metric and symmetry as per the α-$Gd_2S_3$-type), no successful solution or refinement was retained. We nonetheless infer that any volumetric difference between possible competing α-$Gd_2S_3$-types and $Sb_2S_3$-types is minimal given the common lattice direction and the √2 relationships between the others (resulting in approximate $Sb_2S_3$ equivalent lattice of $a$ = 3.016 Å, $b$ = 7.900 Å, $c$ = 7.914 Å, or less than 1% from that in the UB matrix above, determined from our observed lattice points). The spontaneous strain for the notional ppv to α-$Gd_2S_3$ transition is estimated at ~0.1 and the average distance that paired atoms must move during the transformation is roughly 1.5Å, nearly two and three times that of the $Sb_2S_3$-type description, respectively. Consequently, a notional α-$Gd_2S_3$-type derivative of the ppv structure has a compatibility, Δ, of 1.734 compared to our solution in $Sb_2S_3$, with a more favourable Δ = 0.473. Similar values are estimated for the transition pv to α-$Gd_2S_3$-type; where the transformation (identity, plus origin shift of ½ 0 ½) , requires significant movement of F and Fe to fulfill our expectation of higher coordinated Fe polyhedra (to 1.6 Å, with average of 1.08Å) and has a compatibility of Δ = 2.505. It can, nonetheless, be expected that in some systems these structures may be competitive; e.g. in $Dy_2S_3$ [40]. The predominance of the α-$Gd_2S_3$ structure in $A_2X_3$ stoichiometries and the twinning shown in figure 6d, suggests that for $ABX_3$ structures to crystallize in this type, a significant reduction in contrasting cation sizes must be achieved, but not at the expense of limiting oxygen-packing in this higher (than $CaIrO_3$) coordinated structure. It is not necessary (by site symmetry) to invoke $B=A$ stoichiometries or site-disorder (i.e. $AAX_3$ or $(A,B)_2X_3$) to crystallize as α-$Gd_2S_3$-type (with $c^5$ symmetry) as all atoms are symmetrically inequivalent. It does make the suggestion more realisable. Alternatively, it perhaps indicates why only $A_2X_3$ stoichiometries have been determined to undergo the $CaIrO_3$- to α-$Gd_2S_3$-type transition, as the twining involved requires that the orthogonal twins produced have coincident anion sites. Therefore, while the cation coordination is different in both sites in the $CaIrO_3$-type $A_2X_3$ compounds, for the α-$Gd_2S_3$-type to subsequently crystallise by a mechanism that generates orthogonal twins the cation-anion distances must have the *potential* to match in both $c$ site anion polyhedra. This is most simply achieved in $A_2X_3$ stoichiometries, where nil contrast the ionic radii would allow this. As the $Sb_2S_3$- and the α-$Gd_2S_3$-type structures are so evidently closely-related but appear to require this distinction, the application of further pressure (and consequent reduction of differences in unlike ionic radii) may lead to an $Sb_2S_3$-to-α-$Gd_2S_3$-type transition at higher loads in $ABX_3$ systems, as per that calculated for $Al_2O_3$ [18].

Calculations have predicted another probable higher pressure phase in the $NaMgF_3$ system [11]. It has been proposed in symmetry $P6_3/mmc$, with NiAs stacking of cations and Na and F filling an $IrAl_3$-like sublattice, with *fdba* Wyckoff symmetry. This proposed model is identical to $InFeO_3$ (Na → Fe, Fe → In) [41] and $InMnO_3$ [42], with the latter unclassified by type. They are, however, equivalent to the $Be_3N_2$ structure-type after an origin shift of (0 0 ½) is applied (to transform *fdba* ↔ *fdca*). Other structures classified as $Be_3N_2$ include $GaInO_3$ [43] (whose components and solid-solutions are also known in bixbyite and corundum structures) and $YAlO_3$ [44] and $YMnO_3$ [45], both with known $GdFeO_3$-type polymorphs. As this probable higher pressure $Be_3N_2$-type and the $CaIrO_3$-type structures share a common subgroup in *Pnma*, it is of value to determine the common transformations between these structures, through our observed $Sb_2S_3$-type. The proposed $Be_3N_2$ structure transforms, using TRANPATH [46], from $P6_3/mmc$, with an index of 6, via matrix (5), to *Pnma* with the F on 4*f* site-splitting to $c^5$. The transformation from ppv to the common subgroup, in *Pnma*, is via matrix (6) with index 2, which in splits the 8*f* F sites to give a near-identical $c^5$ subgroup structure (Fig. 7).

$$\begin{pmatrix} 0 & 1 & 1 & \vdots & 1/2 \\ 0 & 0 & 2 & \vdots & 1/2 \\ 1 & 0 & 0 & \vdots & 0 \end{pmatrix} \qquad P6_3/mmc \rightarrow Pnma \qquad (5)$$

$$\begin{pmatrix} 0 & 1 & 0 & \vdots & 3/4 \\ 0 & 0 & 1 & \vdots & 3/4 \\ 1 & 0 & 0 & \vdots & 1/2 \end{pmatrix} \qquad Cmcm \rightarrow Pnma \qquad (6)$$

It is quite apparent that this transition path between our experimentally determined ppv structure and our assignment of the $Be_3N_2$-type classification to the $P6_3/mmc$ description of ref. [11] results in the closely-related subgroups S1 and S2 (Fig. 7). These subgroups, in $Pnma$ bear strong resemblance to our experimentally derived $Sb_2S_3$. This reinforces the higher pressure description proposed, even in spite of the substantial lattice compression and distortion required for this transition. We estimate, from symmetry and bond length considerations, that this transformation may occur at about 55 % total compression, per formula unit, from ambient ($a \sim 3.34$ Å, $c \sim 6.25$ Å, $Z = 2$).

**Discussion**

The use of high-pressure X-ray techniques has allowed the study of a series of phase transitions in $NaFeF_3$, leading to a detailed in-situ structural characterisation of pv, ppv and $Sb_2S_3$-type post-post-perovskite phases. This $Sb_2S_3$-type post-ppv structure (a continuously variable structure-type that can be subcategorised into $Th_2S_3$-, $Sb_2S_3$- and $U_2S_3$-types, depending on lattice metrics) has been calculated and observed as high density form of $Al_2O_3$ [18] and $Ti_2O_3$ [22]. Alternatively, in other $A_2X_3$ systems (e.g. $In_2O_3$, $Sc_2O_3$), the α-$Gd_2S_3$-type is preferred and a new systematic was proposed for high density C-type sesquioxides based on this; and further differentiated, in appropriate systems, by the effect of *d*-electrons on related observations [21]. The $Sb_2S_3$-type structure was also proposed as a possible model for a post-ppv form in $NaMgF_3$ [11]. It has however not been observed as a post-$CaIrO_3$-type in either $NaMgF_3$ (or any other fluoride) nor as a post-$CaIrO_3$-type sesquioxide. The lack of any such observation and the interpretation of soft phonon behaviour in $NaMgF_3$ [11] (not the case for the identical transition in $Al_2O_3$ [18] and, unfortunately, the transition is not described from calculations in $NaFeF_3$), have lead to the suggestion that the ppv-to-$Sb_2S_3$ transition is metastable with respect to decomposition to NaF and cotunnite-type $MgF_2$ [11]. However, according to calculations and the conclusion of [11], the enthalpy difference between the dissociation is sufficiently low to possibly allow overstepping decomposition by as little as 2.5 GPa (in $NaMgF_3$) and observing the $Sb_2S_3$-type structure directly. Similarly, the expected small energetic differences between competitive fluoride perovskites are apparently sufficient to allow for an exploration of the complete high pressure, temperature phase diagram of pv and ppv forms of $NaNiF_3$ without any impediment from decomposition, which plagues the same exploration in $NaCoF_3$ [6, 9, 10]. In the absence of any estimation of energetics of the system, other indicators can be used to assess the likelihood of the structures being otherwise compatible with the topology of the phase diagram of $Al_2O_3$, $NaMgF_3$, and other sulphide examples. For comparison, the ratio of the volumes of the polyhedral units to that of the formula unit, $(V_{Na}+V_{Fe})/V_{NaFeF3}$, observed for the ppv form (0.61) and the $Sb_2S_3$-form (0.67) favour very well with those determined for $Al_2O_3$ in $CaIrO_3$-(0.62) and $U_2S_3$-types (0.69) [18], as well as $NaMgF_3$ (0.63, 0.67)[11] and sulphide analogs (see 18). Unsurprisingly, given the relationships between $Sb_2S_3$, $\alpha$-$Gd_2S_3$ and $CaIrO_3$ the value for the notional $\alpha$-$Gd_2S_3$ structure (with metrics produced by the twinned lattice above) is 0.68; while that estimated for the $Be_3N_2$-type is ~1.00. Therefore, with all other indicators being favourable, it is conceivable that changing system would provide, especially at low temperature (as increased temperature favours decomposition) a means to exert chemical pressure to overstep this reaction and transform directly from ppv to $Sb_2S_3$-type. Once the transition occurs, further decomposition can be precluded, as suggested by the phase diagram of $NaMgF_3$ [11] and the highly vertical transition expected for $CaIrO_3$-to-$Sb_2S_3$-type in $Al_2O_3$ [18]. Such advantageous use of small energetic differences may, in the future, permit the study of the higher pressure transition to $Be_3N_2$-type (which is not soft mode) to be investigated, at >200 GPa [11], if, at increased density, packing and polyhedral volumes are commensurate with predictions. While it is evident that $Al_2O_3$ has no such decomposition, the ppv-to-$Sb_2S_3$ and $Sb_2S_3$-to-$\alpha$-$Gd_2S_3$ transitions are expected at pressures exceeding 370 GPa [18] and are also, consequently, beyond all static experimentation. Therefore, for any experimental investigation of this transition to happen, the decomposition (if it is relevant) must be overcome and for the study of the mechanism (etc) of these transitions, suitable analogues must be chosen. This calls for further judicious use of single-crystals, as it is not given that solid-compressed (for laser-heating or large-volume devices) powder samples could be annealed, to recover a strain-free sample with suitable quality data, while still avoiding decomposition. We demonstrate here that, with this combination of technique and sample, that such a course of action is possible with single-crystal $NaFeF_3$. Our observation that the single-crystal-to-single-crystal transition, which back-transforms to the ppv form with little observable hysteresis, taken with the fact that the martensitic transitions proposed should offer no obvious kinetic impediment to the transformation from $CaIrO_3$-type to the $Sb_2S_3$-type. This then clearly generates a further point of alignment between sesquioxide- and perovskite-like systems; through calculation for $Al_2O_3$ [18]. This lends further weight to the proposition that 'multi-megabar crystal chemistry of planet-forming minerals might be related to the rare-earth oxides' [18].

Our description of pv and ppv structures, is also further evidence of the continued development of inherited texture during the transformation process from pv to ppv, as previously discussed elsewhere [14,15,38]. This effect is extended into the $Sb_2S_3$-type structure, and can be pushed further using our descriptions of the transformations to $Be_2N_3$- or α-$Gd_2S_3$-types to assess the structure-property relationships over a wider range of proposed $A_2X_3$-$ABX_3$ topologies. Regarding the ppv-to-$Sb_2S_3$ transition; as the crystallographic orientations are preserved, the low values for both atomic displacement and spontaneous strain, coupled with a more equant $Sb_2S_3$-type lattice, imply that anisotropy will be maintained, at a reduced level. Studies that are capable of extracting the elastic constants of this material can elucidate this further and allow for derivation of rheological properties while making use of the relative geometries, and associated texture development with preferred orientation of these subsequent crystal structures. One other aspect of the topotactic relationships discussed above is that the $CaIrO_3$-type forms part of a homologous series of structure types that includes spinels, post-spinels and oxides [47-49] and these have been demonstrated to be capable of forming intergrowths with other members of the series, and with rocksalt structures [50]. The importance of such an assemblage is obvious given that $(Mg,Fe)_2SiO_4$ will decompose towards pv + rocksalt prior to the transition to ppv. The other salient point that arises from inspection of this series is that all more complex forms *tend* towards topologies based on decoration of a sesquioxide- or ppv-like structural basis. Upon increasing coordination, and forming the $Sb_2S_3$-type structure, these $A_2X_3$ and $ABX_3$ can then continue in an analogous manner, with only a minor effect on the solubility or capacity to be intergrown, within this series of stoichiometries. This generalism may indeed extend further than previously supposed, and beyond structures limited by 6-fold coordinated cations, in line with arguments for other sesquioxides [18], given the evident parallels between the $Be_3N_2$ structure (most easily viewed along the <110> direction; Fig. 2c of ref. [11]) and, e.g. Figs. 3 and 5 of ref. [47]. Similar comments are made regarding possible intermixing of ppv and PPV-II types elsewhere [28].

In conclusion, we provide a full in-situ structural description for all observed structures in the $NaFeF_3$ system up to a post-post-perovskite phase. This $Sb_2S_3$-type phase is, to the best we can determine via single-crystal methods at RT, stable. Whether the occurrence of the $Sb_2S_3$-type structure in the fluoride system truly reflects silicate oxide systems is beyond all experimentation; as the transition is expected at 1.6 TPa [51]. Nonetheless, the topotactic relationship between this and the lower density structures is maintained and can be extended, using our crystallographic orientations (UB or transformation matrices) combined with calculated tensor information to describe the rheological response of the observed $Sb_2S_3$ structure. Our experimental evidence is aligned with the results of calculations regarding the mechanism for such a transition [11]. Similar transformation matrices provide for transitions to and from the α-$Gd_2S_3$ and $Be_3N_2$-type models, which permits further such investigation and modeling, built upon a strict crystallographic basis, particularly in sesquioxide and double-sesquioxide perovskites, where these assemblages are more likely. The observation of the $Sb_2S_3$ and our description of a possible mechanism, via orthogonal twinning of a post-$CaIrO_3$-type to form the α-$Gd_2S_3$ structure, provides a link between the generalized transition sequences of $A_2X_3$ and $ABX_3$ structures that have been observed to crystallize in the ppv structure under extreme conditions. With the full suite of matrix and twinning descriptions included, we promote that the full range of pv, ppv, $Sb_2S_3$, α-$Gd_2S_3$ and $Be_3N_2$-type structures can be then reproduced from any other starting point; given that the perovskite to $Rh_2O_3$(II) relationship is known already. We also offer reasoning for why the α-$Gd_2S_3$ structure may be preferred over the $Sb_2S_3$-type in certain $A_2X_3$ compounds. Finally, given that $CaIrO_3$- $Sb_2S_3$- and α-$Gd_2S_3$-types can all be related through twinning and matrix descriptions that involve little or no volume change, we expect that any constraints from the structures themselves on the solubility and continued transfer of elements upon transition to the post-post-perovskite forms to be minimal. Such a transformation might be expected, then, without any significant loss of $Al^{3+}$, $Ti^{3+}$ or $Fe^{3+}$ content in more complex chemistries (and we note that these are all trivalent). Ppv can transform to its denser $Sb_2S_3$-type polymorph at higher pressures and may be competitive with α-$Gd_2S_3$ over an extended range of pressures. Any transition to a higher pressure phase will preclude any retrograde transformation at higher temperatures, and/or decomposition to its components. The derivation of a common subgroup between our observed ppv and $Be_3N_2$-type $NaFeF_3$ that closely matches our observed data for $Sb_2S_3$ provides further justification for the use of this structural sequence in modeling higher density environments in any related system. Indeed, the only catalogued other $Be_3N_2$-type structures are also known in corundum-, perovskite- and bixbyite-type structures, all commonly occurring as ante-ppv or ante-α-$Gd_2S_3$ structures in $A_2X_3$ systems. These descriptions can be generalized for further exploration, by experiment or otherwise, for both $ABX_3$ and $A_2X_3$ systems, and using the inherited directional properties that come with UB descriptions, further tested against expected, e.g. anisotropic signatures.

The search for similar experimental descriptions of the higher pressure transitions of perovskites towards possible α-$Gd_2S_3$- and/or $Be_3N_2$-types is, at in excess of 210 GPa in the nearest calculated, $NaMgF_3$ chemistry [11], beyond current experimental technology in single-crystal measurements. However, as we highlight in this work, the use of analogous compounds is of great utility in experimental exploration of the topology of generalized phase diagrams and transitions. Suggestions, from this work, for such studies are to use those compounds, of $ABX_3$ stoichiometry, with components found in $A_2X_3$ systems i.e. double sesquioxides where $ABX_3 = A^{3+}{}_2B^{3+}{}_2X_6$. In other words, these high density structures, some of which may be precluded in silicates by decomposition, are certainly possible in other sesquioxide-based compounds of an analogous $ABX_3$ stoichiometry.

The ESRF is thanked for allocation of in-house beamtime at ID06LVP. The studentship of FLMB is supported by NFR project 214260 of the Norges forskningsråd to SM.

**Table 1**

**Results of single-crystal refinement from data collected at 9.5 GPa, GdFeO$_3$-type NaFeF$_3$:**

| Compound | NaFeF$_3$-I | | | | | | |
|---|---|---|---|---|---|---|---|
| Space group | *P n m a* | 62 | | | | | |
| | | | | | | | |
| Lattice parameters, Å | | | | | | | |
| | *a* | *b* | *c* | | | | |
| | 5.572(6) | 7.4607(8) | 5.119(4) | | | | |
| | | | | | | | |
| Unit-cell volume | | 212.802 Å$^3$ | | | | | |
| | | | | | | | |
| Atom | X | Y | Z | Occ | U Å2 | Site | Sym |
| Na | 0.0760(6) | 0.25 | -0.024(3) | 1 | 0.0171(9) | 4c | .m. |
| Fe | 0.5 | 0.0 | 0.0 | 1 | 0.0120(4) | 4b | -1 |
| F1 | 0.4451(9) | 0.25 | 0.135(3) | 1 | 0.0158(16) | 4c | .m. |
| F2 | 0.3092(7) | 0.0615(7) | -0.328(2) | 1 | 0.0189(12) | 8d | 1 |
| | | | | | | | |
| Selected distances | | | | | | | |
| 2x | Fe1-F2 | 1.989(7) Å | Na1-F1 | 2.121(18) Å | | Na1-F2 | 2.183(8) Å |
| 2x | Fe1-F1 | 2.013(5) Å | Na1-F1 | 2.213(10) Å | | Na1-F2 | 2.465(11) Å |
| 2x | Fe1-F2 | 2.038(9) Å | | | | | |

**Table footnote:** 36% completeness to 0.61 Å, 50% to 0.8 Å. Rint 5.7 % and 5.2% over the same ranges, from 465/225 and 347/145 measured/unique reflections. Refined to Robs = 3.67%, wRobs = 4.31%, wRall 4.31%.

**Table 2.**

**Single-crystal refinement results from NaFeF₃ ppv, from data taken at 12.0 GPa on decompression.**

| Compound | NaFeF$_3$-II | | | | | | |
|---|---|---|---|---|---|---|---|
| Space group | *C m c m* | 63 | | | | | |
| Lattice parameters, Å | | | | | | | |
| | *a* | *b* | *c* | | | | |
| | 2.9731(14) | 9.453(12) | 7.120(3) | | | | |
| Unit-cell volume | | 200.106 Å$^3$ | | | | | |
| Atom | X | Y | Z | Occ | U Å2 | Site | Sym |
| Na | 0.0 | 0.229(7) | 0.25 | 1 | 0.003(5) | 4c | 2/m.. |
| Fe | 0.0 | 0.0 | 0.0 | 1 | 0.019(5) | 4a | m2m |
| F1 | 0.0 | 0.079(8) | -0.25 | 1 | -0.006(6) | 4c | m2m |
| F2 | 0.5 | -0.162(5) | -0.069(2) | 1 | -0.010(5) | 8f | m.. |
| Selected distances | | | | | | | |
| 4x | Fe1-F2 | 1.93(3) Å | 4x | Na1-F2 | 2.11(3) Å | | |
| 2x | Fe1-F1 | 2.018(3) Å | 2x | Na1-F1 | 2.35(8) Å | | |

**Table Footnote:** Rint obs/all 10.63/10.66 from 48/53 averaged from 93/107 reflections extracted from integration in *oC* lattice. Four reflections culled at 20% difference from average. Refined on F, with 10*σFobs cutoff (19 reflections). Refines to Robs 16.48 wRobs 11%, wRall = 11.4%.

**Table 3**

**Single-crystal refinement results from the post-post-perovskite, $Sb_2S_3$-type, $NaFeF_3$ at 20.5 GPa, from data collected upon decompression.**

| Compound | $NaFeF_3$-III | | | | | | |
|---|---|---|---|---|---|---|---|
| Space group | *P n m a* | 62 | | | | | |
| Lattice parameters, Å | | | | | | | |
| | *a* | *b* | *c* | | | | |
| | 7.8380(14) | 3.007(3) | 7.901(5) | | | | |
| Unit-cell volume | | 186.218 Å$^3$ | | | | | |
| Atom | X | Y | Z | Occ | U Å2 | Site | Sym |
| Na | 0.326(2) | 0.25 | -0.024(4) | 1 | 0.024(4) | 4c | .m. |
| Fe | -0.0080(8) | 0.25 | 0.2047(10) | 1 | 0.010(2) | 4c | .m. |
| F1 | 0.2191(17) | 0.25 | 0.311(4) | 1 | -0.006(4) | 4c | .m. |
| F2 | -0.061(2) | 0.75 | 0.371(5) | 1 | 0.013(5) | 4c | .m. |
| F3 | -0.3794(18) | 0.75 | 0.424(4) | 1 | -0.003(4) | 4c | .m. |
| Selected distances | | | | | | | |
| 1x | Fe1-F1 | 1.967(19) Å | 1x | Na1-F3 | 2.01(3) Å | | |
| 2x | Fe1-F3 | 2.04(2) Å | 2x | Na1-F1 | 2.02(3) Å | | |
| 2x | Fe1-F2 | 2.076(16) Å | 2x | Na1-F3 | 2.12(3) Å | | |
| 1x | Fe1-F1 | 2.142(15) Å | 2x | Na1-F2 | 2.34(2) Å | | |
| 1x | Fe1-F2 | 2.386(2) Å | | | | | |

**Table Footnote:** 147/348 unique reflections. Rint obs/all 12.25/13.14 from 91 reflections averaged from 146/326. Three reflections culled at 20% difference from average. Refined on F with I < 3σI set as unobserved and 15 *σFobs rejection (15 reflections). Refines to Robs 16.7%, wRobs 11.8% and wR(all) of 13.57%.

**Figure Legends**

**Figure 1. The structures of NaFeF$_3$.** The observed structures of NaFeF$_3$ described in the text and resulting from refinements in Tables 1-3. The Na-atoms and polyhedra are yellow, Fe polyhedra are brown in the (a) GdFeO$_3$-type perovskite, (b) CaIrO$_3$-type ppv and (c) Sb$_2$S$_3$-type structures. Annotations are referred to in the text. The figures have been prepared (as Figs 4, 5, 7) using VESTA [52].

**Figure 2. Powder diffraction data.** Rietveld refinements of high-pressure x-ray patterns of NaFeF$_3$ taken from large-volume samples at about 9.5 GPa, before (a) in the pv structure, and after transition (b) in the ppv structure. Time-lapsed powder diffraction data are also shown as a stacked waterfall plot (c), taken during compression to 9.5 GPa, to highlight the increase in anisotropy with pressure, which increases upwards.

**Figure 3. Experimental electron density distributions.** The Fobs maps for the single-crystal specimens calculated in Jana2006 [53] of (a) ppv and (b) Sb$_2$S$_3$-type NaFeF$_3$. These distributions highlight the increased bond distances and the new staggering of the Fe-sites. They are directly comparable to those presented in ref [11], at 90°.

**Figure 4. Schematic structural relations between ppv and Sb$_2$S$_3$.** Panels (a)-(d) schematically represent the transformation, derived using STRUCTURE RELATIONS [35], for a transition between refined ppv (a) and Sb$_2$S$_3$ (d) structures with parameters here identical to those presented in Tables 2 and 3. Panels (b) and (c) are shown flattened and illustrate the constructions obtained when the high symmetry (*Cmcm*) ppv is transformed into the low symmetry *Pnma* lattice of the Sb$_2$S$_3$ structure with ppv dimensions (b). In (c) the Sb$_2$S$_3$ structure is lattice distorted to this same metric. The distances the atoms must move and the spontaneous strain for this transformation can be effectively estimated from (b) → (c) and (c) → (d), as discussed in the text.

**Figure 5. Atom displacements in the ppv to Sb$_2$S$_3$ structure.** The atom positions of the Sb$_2$S$_3$ structure have been superposed upon the *b*-axis projection of ppv structure to highlight the loss of symmetry and for direct comparison with unstable mode calculations shown in Fig. 4 of ref [11]. High symmetry axis directions and planes are indicated by dashed lines, along which both the Na and Fe are located in ppv. They are displaced from these positions upon transition to Sb$_2$S$_3$. Atom movement is indicated by red arrows, so that Na atoms move from yellow to blue positions and F from grey to green. The Fe positions are occluded by the polyhedra.

**Figure 6. The relative orientations of observed lattices.** Experimentally harvested lattice points and indexed lattices from high pressure single-crystal runs on NaFeF$_3$. (a) shows the pv lattice points, (b) shows the Sb$_2$S$_3$ lattice points. In each case these are overlain by the ppv lattices. (c) Shows a composite reconstruction of the three lattice directions, for pv (black), ppv (blue) and Sb$_2$S$_3$(red) lattices, referred to by matrices (2)-(3). Panel (d) demonstrates that a twinned pair of orthogonal α-Gd$_2$S$_3$-type lattices can index (but not refine) Sb$_2$S$_3$ reflection data. The marked angles are referred to in the text.

**Figure 7. Schematic transition paths between post-ppv structures.** Clockwise from left: The experimentally determined ppv structure (Table 2) transforms to the Sb$_2$S$_3$ structure (Table 3) upon application of pressure. The related chemistry, NaMgF$_3$, is proposed [11] to undergo a transition to a structure here classified as Be$_3$N$_2$-type. Using TRANPATH [46], we can investigate common subgroups of the two high symmetry ppv and Be$_3$N$_2$ structures, through their expression in the common subgroup symmetry. It is remarkable that these subgroup intermediates, S1 and S2, coincide rather closely with each other and with the experimentally determined Sb$_2$S$_3$-type NaFeF$_3$.

**Figure 1.**

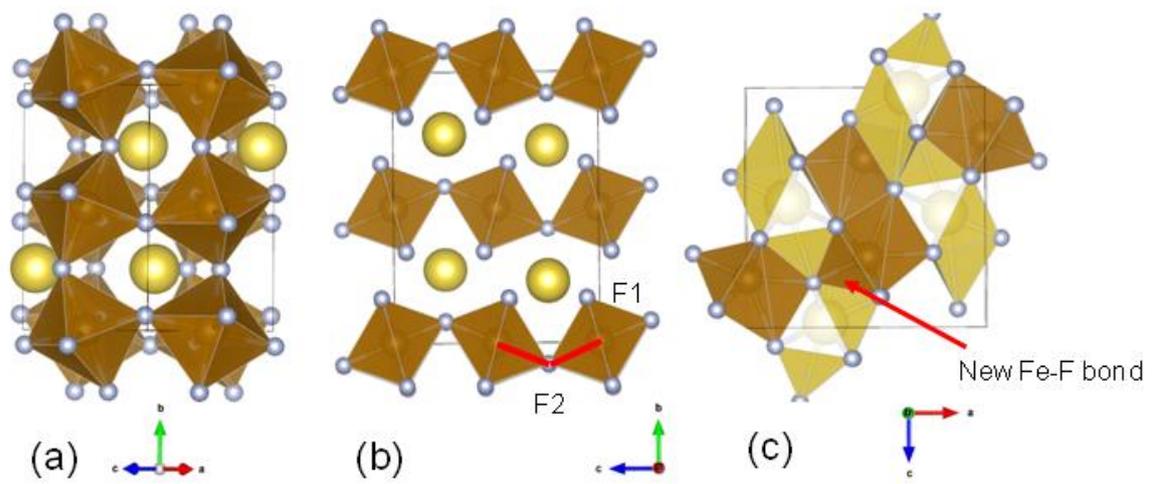

**Figure 2.**

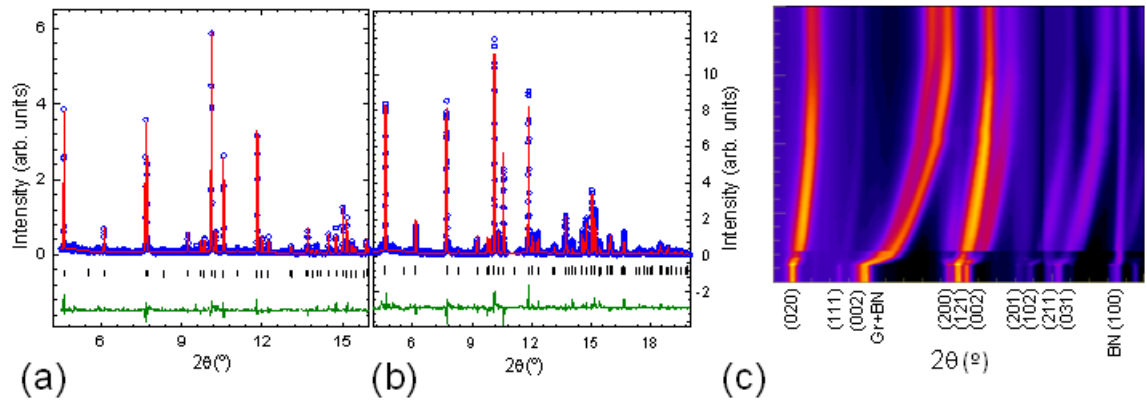

**Figure 3.**

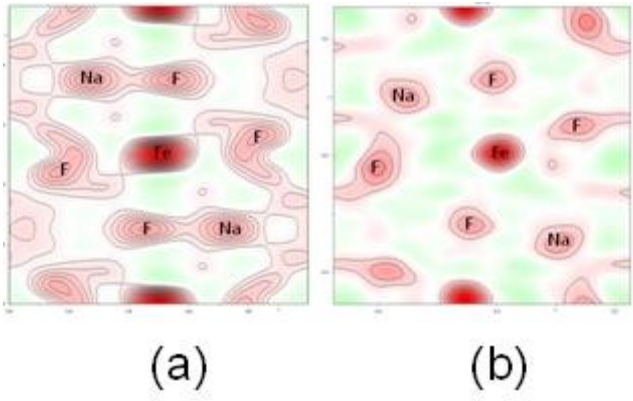

**Figure 4.**

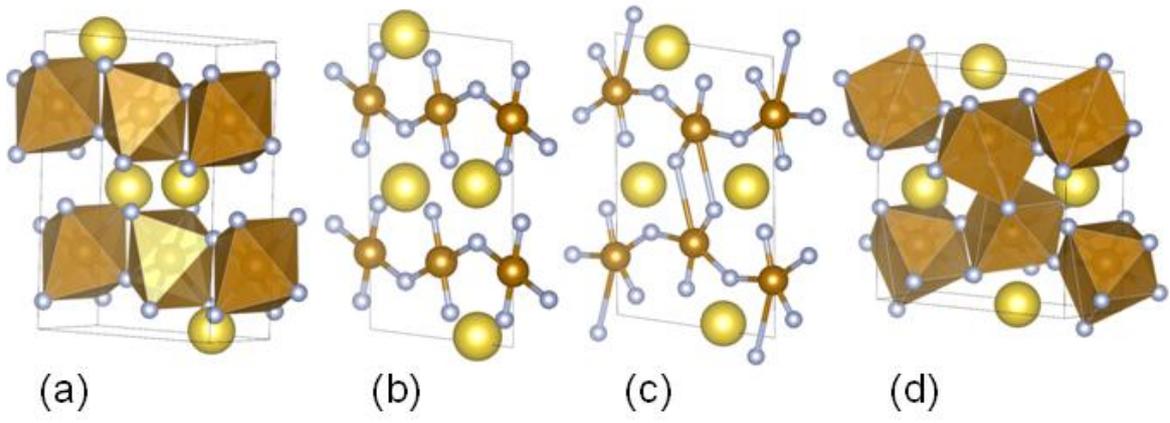

(a) (b) (c) (d)

**Figure 5.**

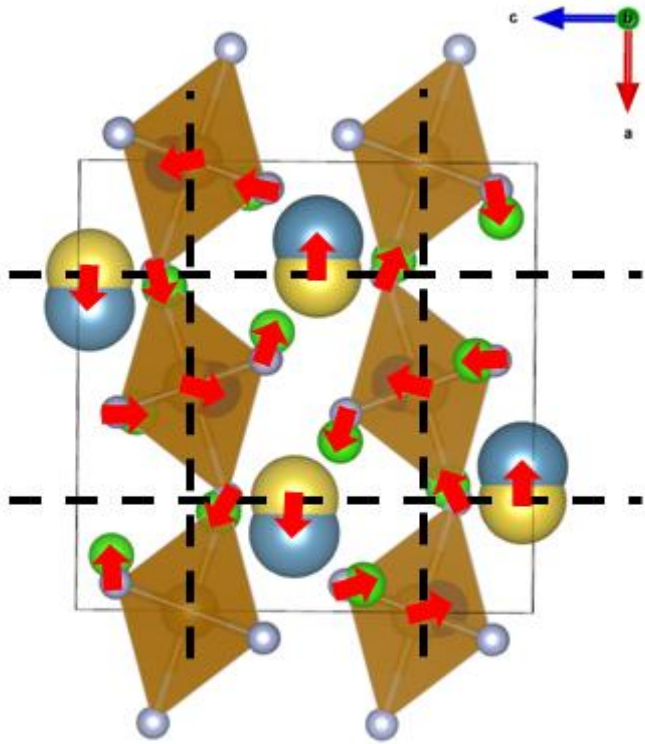

**Figure 6.**

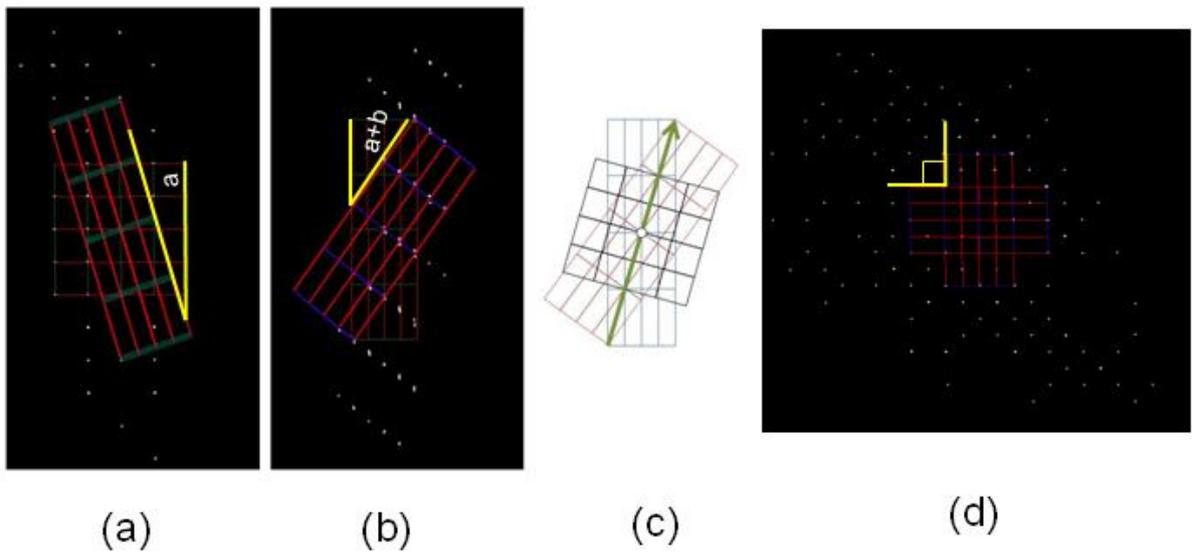

(a)　(b)　(c)　(d)

**Figure 7.**

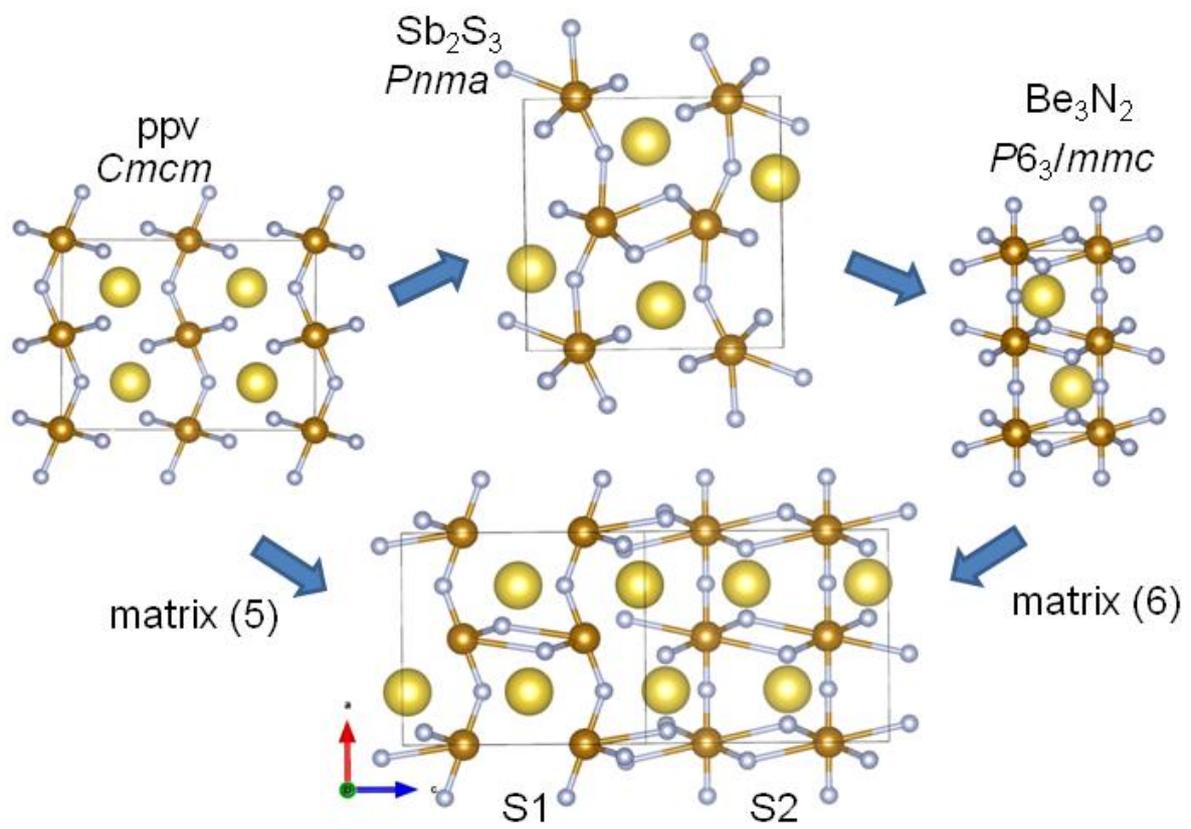

# Supplementary Information

## Observation of the $Sb_2S_3$-type post-post-$GdFeO_3$-perovskite: A model structure for high density $ABX_3$ and $A_2X_3$ phases.

Wilson A. Crichton 1*, Fabian L. M. Bernal 2, Jérémy Guignard 1, Michael Hanfland 1 and Serena Margadonna 2*

1. ESRF - The European Synchrotron, 71, rue des Martyrs, Grenoble cedex, 38043, France.

2. Department of Chemistry and Center for Materials Science and Nanotechnology, University of Oslo, Oslo, Norway.

## Materials and Methods

**Synthesis**

The $NaFeF_3$ samples were synthesized according to ref. 1 and 2. NaF, Fe and $FeF_3$ powders were finely ground and mixed in a 3:2:1 ratio. The so-obtained mixtures were then pelletized and loaded into a Fe-tube which was used as a reaction vessel. All the manipulations were carried out in an Ar-filled glove-box given the hygroscopic nature of NaF and to avoid exposure to oxygen and/or moisture. The Fe-tube was tightly closed and then heated under a constant nitrogen flow at 750 ºC for 8 days.

**Single-crystal diamond-anvil cell techniques**

Single-crystal samples were loaded in a membrane-driven diamond-anvil cell, together with a ruby sphere for pressure measurement and a He pressure-transmitting medium. Single-crystal datasets were collected at a wavelength of 0.41458Å, selected by a horizontally focusing Laue monochromator at beamline ID09A of the ESRF, by slicing omega from -30º to 30º with an integrated interval of 0.5º per frame [3]. The exposure time per frame was 1 second and the beam size was approximately 0.01 x 0.01 mm. The sample-detector distance and wavelength were refined using a Si powder sample, following the method described in ref. [4]. Data were reduced using the CrysAlis software of Oxford Diffraction. In general, the model system (laboratory and crystal orientation) was refined with sample-detector distance fixed at that determined by Fit2D. The data reduction followed averaging in the centred lattice of the eventual space group assignment and the frame data were rescaled using ABSPACK correction for absorption. The data were refined using Jana2006 [5], following averaging and culling of >20% I/Ī for equivalents. Further corrections to the assembled *hkl* dataset were made on an individual basis and were dependent on lattice and averaged Laue symmetry, degree of overlap, interception of peaks with diamond reflections, detector edges etc.

**Powder diffraction from large-volume press**

Large-volume press data were recorded at beamline ID06LVP, ESRF, using a linear pixelated GOS detector from Detection Technology, running sequential exposures for 3.2 seconds at 10 Hz at 32 seconds interval and mounted to intercept the downstream diffraction from the horizontal anvil gap at 1180 mm distance. The detector-beam normal plane was mechanically corrected for tilt and rotation and the detector position for zero-offset and calibrated against SRM660a, using Fit2D [4], at monochromatic wavelengths of 03757 Å or 0.2254 Å, selected from the emission of a U18 undulator (6.1-6.2 mm gap) by a Cinel Si(111) double-crystal monochromator. Data for refinement were collected using a Tl:NaI scintillator counter, with a 0.2 mm pinhole and receiving slits at intermediate distance, which acted as collimators of background scattering coming from the cell assembly. The effective radius of the pseudo-rotation detector arm is 1350 mm. Collimated beam sizes were 0.5 mm horizontal by 1 mm, or the vertical anvil gap, if smaller. Data were refined using GSAS [6,7]. In each case the NaFeF$_3$ sample was finely ground in a corundum mortar and loaded into a hBN (Goodfellow) capsule, before being included into the 10/4 windowed Cr:MgO assembly. Pressure was generated using the 2000 ton MAVO press in 6/8(x32) mode, with carbide anvils [8]. Pressures were estimated using the equation of state of h-BN [9].